\documentclass[preprint,showpacs]{revtex4}
\usepackage{graphicx}

\begin{document}

\title{Two dimensional scaling  of resistance in  flux flow region in $Tl_2Ba_2CaCu_2O_8$ thin films}

\author{X.F. Lu$^{1}$}
\author{Z. Wang$^{1}$}
\author{H. Gao$^{1}$}
\author{L. Shan$^{1}$}
\author{Y.Z. Zhang$^{1}$}
\author{R.T. Lu$^{2}$}
\author{L. Fan$^{2}$}
\author{S.L. Yan$^{2}$}
\author{Z.D. Wang$^{3}$}
\author{H.H. Wen$^{1}$}
\author{H. H. Wen}\email{hhwen@aphy.iphy.ac.cn}

\address{$^1$National Laboratory for Superconductivity, Institute
of Physics, Chinese Academy of Sciences, P.~O.~Box 603, Beijing
100080, P.~R.~China}
\address{$^2$Department of Electronics,
Nankai University, Tianjin 300071, P.~R.~China}
\address{$^3$Department of Physics, University of Hong Kong, Pokfulam Road,
Hong Kong, P.~R.~ China}

\begin{abstract}
The resistance of $Tl_2Ba_2CaCu_2O_8$ thin films has been measured
when the angle between the applied fields and $ab$-plane of the
film is changed continuously at various temperatures. Under
various magnetic  fields, the resistance can be well scaled in
terms of the $c$-axis component of the applied fields at the same
temperature in the whole angle range. Meanwhile, we show that the
measurement of resistance in this way is a complementary method to
determine the growth orientation of the anisotropic high-$T_c$
superconductors.
\end{abstract}


\maketitle

\section{Introduction}
The research on vortex dynamics in high-$T_c$ superconductors (
HTS ) is important  since it has close connection with the
potential  applications  in the  future.  It is known that the
vortex system of the strong anisotropic cuprate superconductors,
such as $Bi_2Sr_2CaCu_2O_8$ or $Tl_2Ba_2CaCu_2O_8$, is very
complicated because of high transition temperature and very short
coherent length, and the vortex dynamics  is particularly rich
when  a tilted field is applied. Many puzzling experimental
phenomena have been observed in the case of tilted vortices  for
high anisotropic superconductors, for example, one-dimensional
chain state of vortex matter observed by Hall probe
microscopy\cite{nature} and Bitter
decoration\cite{BitterDecoration}, stepwise behavior of
vortex-lattice melting transition in tilted magnetic
fields\cite{Stepwise}, suppression of irreversible magnetization
by in-plane field\cite{Supression,Supression2}, etc. The vortex
dynamics has been studied extensively\cite{Wen-prl1,Wen-prl2} when
the magnetic field is applied along $c$-axis of
$Tl_2Ba_2CaCu_2O_8$, however, the combined vortex lattice phases
still remain largely unexplored in the tilted vortices scenario.

In this paper, we present the resistance of two
$Tl_2Ba_2CaCu_2O_8$ thin film samples in flux flow region under
tilted magnetic fields  when the angle between the applied fields
and the $ab$-plane of the thin film is changed continuously. One
of the measured thin film is grown on $<001>$ $SrTiO_3$ ( $STO$ )
substrate, and another is grown on a tilted $STO$ substrate with
the surface plane   about 20$^\circ$ away from the $<001>$ plane.
The resistance as a function of  angle demonstrates
self-consistency. Moreover, the resistance at the same fixed
temperature can be well scaled  versus the $c$-axis component of
the applied magnetic fields. In addition, the measurement of
resistance in this way may offer a complementary method to
determine the growth orientation of the highly anisotropic
superconductors, and it is a very convenient method to identify
the tilt angle of the tilted-growth thin films.

\section{Experiment Details}
The $Tl_2Ba_2CaCu_2O_8$ films were deposited onto $STO$ substrates
using the dc magnetron sputtering method. One of which was
$c$-axis-oriented film ( denoted as $S$-$c$ ), and another  (
denoted as $S$-$t$ ) was deposited on the substrate with surface
cut at an angle about 20$^\circ$ with respect to the $<001>$ basal
plane( see Fig. 1 (C) ). The details about the preparation  of
samples can be found in other
literatures\cite{yan-apl,Tilted-20D}. The thickness of the thin
film is about 3000 \AA. The size of $S$-$c$ is about 5$\times$0.5
$mm^2$, and $S$-$t$ is patterned into a micro-bridge of 3 $um$ in
length and 3 $um$   in width using photolithography and ion beam
etching techniques. The resistance measurement was done with
standard four-probe technique. In the case of sample $S$-$t$, the
gold leads were attached by silver paste to the gold pads
deposited onto the surface of the film. The contact resistance
between the electrodes and sample was below 1 $\Omega$. The
resistive transition curves are shown in Fig. 2  and the  onset
transition temperature ${T}_{c}^{onset}$ is about 120 $K$ for both
samples.

\begin{figure}
       \centering
       \includegraphics[width=7.5cm] {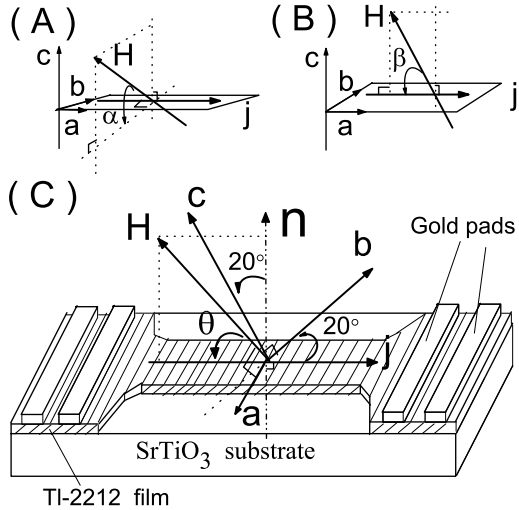}
       \caption{The configuration of field, transport current and the  schematic structure of the sample ( not to scale ):
       (A), (B) for thin film $S$-$c$ with angle $\alpha$ and $\beta$,
       (C) for tilted thin film $S$-$t$, $\vec{n}$ is the normal direction of substrate surface. }
       \label{figure1}
       \end{figure}

\begin{figure}
       \centering
       \includegraphics[width=7cm] {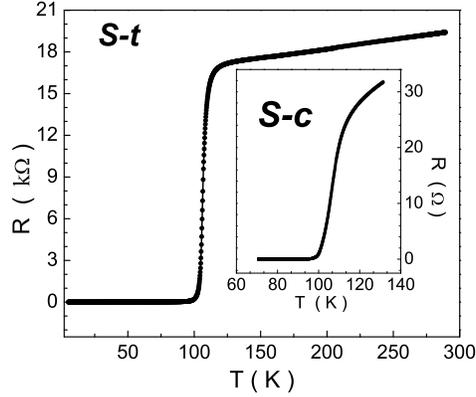}
       \caption{The resistive  transition curves  under zero field of $Tl$-2212 thin films( $R$ $vs$ $T$ ).
               The inset for $S$-$c$ thin film.}
       \label{figure2}
       \end{figure}

The measurement of resistance  was carried out on an Oxford
cryogenic MagLab system ( MagLab12Exa ) with  magnetic field up to
12 Tesla and temperature ranging from 1.5 $K$ to 400 $K$. The
resistance was measured by bipolar mode, so each data point was
the average of the measured results for the positive and negative
transport current. Because the sample platform can be rotated from
0$^\circ$ to 360$^\circ$ by a controlled  step motor during the
measurement, the angle between the sample and the magnetic field
can be changed from 0$^\circ$ to 360 $^\circ$ continuously. So the
$ab$-plane ( or $c$-axis ) of the film can be coordinated exactly
by the minimum ( or maximum ) of the resistance during the
rotation of the film under applied field at fixed temperatures in
the mixed state. The magnitude of the field and its direction can
be varied precisely ( error bar 0.1$^\circ$ ) and freely, which is
particularly helpful and indispensable for the measurement of the
tilted film. In the measurement process, the dc current I, goes
through the film along the surface of the film, and the angle
between the applied field and the film surface $\alpha$ ( or
$\beta$, $\theta$ ) is changed continuously ( see Fig. 1 ). Here,
$\alpha$ is defined by the angle between the field and the
$ab$-plane when $\vec{H}$ is always perpendicular to the $\vec{j}$
( Fig.1 (A) ), $\beta$ is the angle between $\vec{H}$ and
$ab$-plane when $\vec{H}$  is always in the  plane of $c$-axis and
$\vec{j}$ ( Fig. 1 (B) ). For the 20$^{\circ}$ tilted film
$S$-$t$, $\beta = \theta+20^{\circ}$ or $\theta-20^{\circ}$, here
$\theta$ is the angle  between $\vec{H}$ and the surface of
substrate ( Fig. 1 (C) ).

\section{Results and Discussions }
For the  film $S$-$c$, we have carefully measured the resistance
under various fields at different fixed temperatures in mixed
state when $\alpha$ ( or $\beta$ ) is changed from 0$^{\circ}$ to
195$^{\circ}$ continuously. For clarity, we only show some
 typical results here. The resistance  under the same
applied field at different temperatures  are shown in  Fig. 3 when
the angle $\alpha$  and  $\beta$  is continuously varied. At the
same temperature,  $R$ as a function of angle $\alpha$ and $\beta$
for various   fields are displayed  in Fig. 4(a) and Fig. 5(a),
respectively. It is clearly found that all these resistive curves
show symmetry about $\alpha$ ( or $\beta$ ) $\approx$
90$^{\circ}$, which indicates that $S$-$c$ is a $c$-axis-oriented
thin film. The resistance as a function of the angle demonstrates
self-consistency in the whole flux flow region. With angle varying
from 0$^{\circ}$ to 180$^{\circ}$, the resistance is changed  from
the minimum to the maximum and then to the minimum in one cycle,
which corresponds to c-axis component of magnetic field changing
from the lowest to highest and then to the lowest.

Moreover, it is found that all curves at the same temperature can
merge to a common one when we draw $R$ as a function of
$H\cdot\sin\alpha$ and $H\cdot\sin\beta$. For example, the scaling
results of  all curves at T = 90 K ( Fig. 4(a) and Fig. 5(a) ) are
shown in  Fig. 4(b) and Fig. 5(b), respectively. In other words,
$R$ is only dependent on  the $c$-axis component of fields, and
this law is also obeyed at other temperatures of mixed state.

According to effective mass model, the resistivity of an
anisotropic superconductor in the mixed state is dominated by the
effective field $H_e$ which can be expressed in the
formula\cite{ZHWang}
\begin{equation}
 H_e=H\times\sqrt{\sin^2\alpha+\cos^2\alpha/\gamma^2}
\end{equation}
here, $\gamma$ is the anisotropy parameter ( $\gamma
 = {(m_c/m_{ab})}^{1/2} = {\lambda}_c/{\lambda}_{ab} =
{\xi}_{ab}/{\xi}_c$ ). Our results of $R$ can be scaled with
${H\cdot\sin\alpha}$ ( or ${H\cdot\sin\beta}$ ) perfectly, the
anisotropy parameter $\gamma$ of $Tl$-2212 must be very large.
This  is consistent with the reported large values of $\gamma$ and
the  nearly two-dimensional limit for Tl-based
system\cite{IYE-comments}.

In order to understand this 2$D$ scaling law from vortex dynamics,
we assume the following  simplified phenomenological  model based
on the fact that the anisotropy parameter $\gamma$ of $Tl$-2212 is
very large. Given a $\vec{H}$, we can simply decompose it into two
parts for the case of angle $\alpha$: $c$-axis component
$H\cdot\sin\alpha$ and $ab$-plane component $H\cdot\cos\alpha$. As
we know, $c$-axis component of field ( $H\cdot\sin\alpha$ )
generates small size pancake vortices ( PVS ) resident in the
$ab$-panes, which causes dissipation due to the easy motion of PVS
driven by in-plane Lorentz force. On the other hand, Josephson
vortices ( JVS )  with density proportional to $H\cdot\cos\alpha$,
are formed with the in-plane component of $\vec{H}$ and subjected
to the $c$-axis Lorentz driving force, however it is much
difficult for the JVS to move across the $ab$-planes  due to the
strong intrinsic pinning of high anisotropic HTS. As a result, the
contribution of JVS to $R$ is negligible compared to that of PVS.
For the case of angle $\beta$, the only $c$-axis-component field
dependence of $R$ is mainly due to the absence of Lorentz force
for Josephson vortex lines, because $\vec{j}$ is parallel to the
in-plane component of field $\vec{H}\cdot\cos\beta$. In fact,
$\vec{H}$ can not be ideally in the plane of  $c$-axis and
$\vec{j}$, and there exits little Lorentz-force along $c$-axis for
JVS. But the strength of the force is negligible compared to the
intrinsic pinning.
\begin{figure}
       \centering
       \includegraphics[width=8cm] {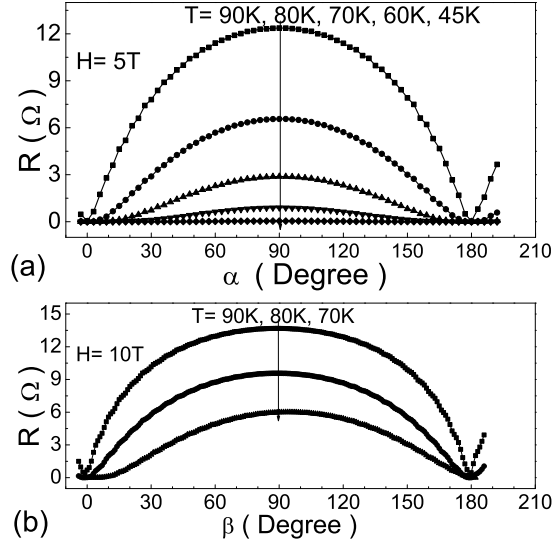}
       \caption{ The field and temperature dependence of resistance as a
       function of angle for  $S$-$c$. (a) R vs angle $\alpha$, T = 90 K, 80 K, 70 K, 60 K, 45 K from the top curve to
       the bottom one  with  H = 5 T. (b)  R vs angle $\beta$, T = 90 K, 80 K, 70 K  from the top  to
        bottom  with  H = 10 T. }
       \label{figure3}
       \end{figure}

\begin{figure}
       \centering
       \includegraphics[width=8cm] {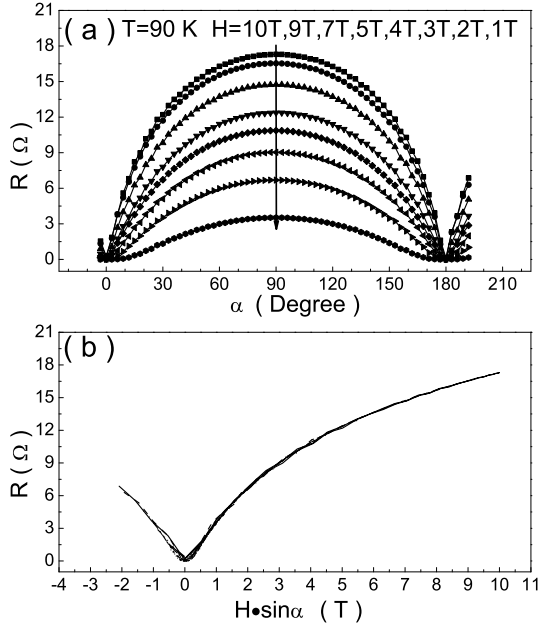}
       \caption{ The angular ( $\alpha$ ) and field dependence of resistance at
       fixed temperature ( $T$ = 90 $K$ ) for $S$-$c$. (a) $R$ vs  $\alpha$,
       H=10 $T$, 9 $T$, 7 $T$, 5 $T$, 4 $T$, 3 $T$, 2 $T$, 1 $T$ from the top to bottom, (b) The scaling of $R$ in
       terms of H$\cdot\sin\alpha$. }
       \label{figure4}
       \end{figure}

\begin{figure}
       \centering
       \includegraphics[width=8cm] {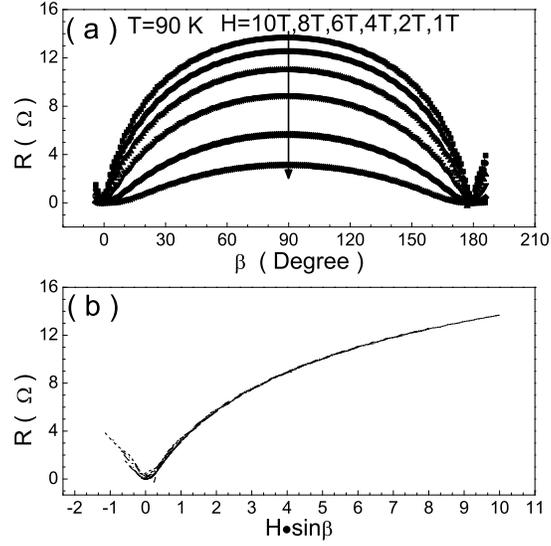}
       \caption{The angular ( $\beta$ ) and field dependence of resistance at
        fixed temperature ( $T$ = 90 $K$ ) for $S$-$c$. (a) $R$ vs  $\beta$,
        (b) The scaling of $R$ in terms of H$\cdot\sin\beta$.[ Because the
        sample was not  patterned and one of voltage electrodes was
        reattached in the four-point method measurement, the  values of
        resistance for $\beta=90^\circ$ are different with  that when
        $\alpha=90^\circ$(Fig. 4) for the same applied field .] }
       \label{figure5}
       \end{figure}

\begin{figure}
       \centering
       \includegraphics[width=8cm] {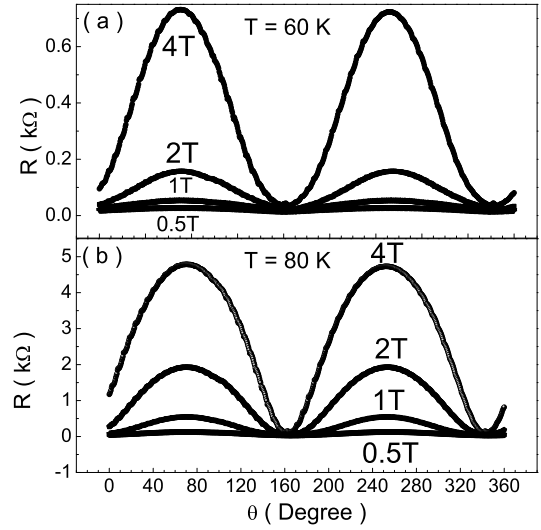}
       \caption{The angular and field dependence of resistance  at
       the fixed temperature  (a) $T$ = 60 $K$  and  (b) $T$ = 80 $K$ for sample $S$-$t$.
       H= 4 $T$, 2 $T$,  1 $T$,  0.5 $T$ from the the top curve to the bottom one in each panel. }
       \label{figure6}
       \end{figure}

\begin{figure}
       \centering
       \includegraphics[width=8cm] {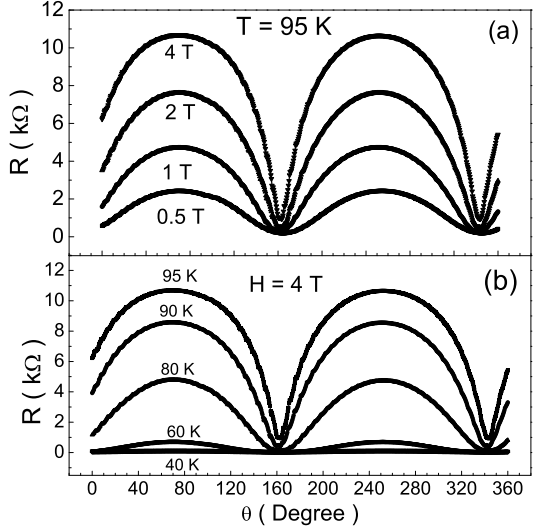}
       \caption{The  field and temperature dependence   of resistance as
       a function of angle $\theta$ for $S$-$t$. (a) H = 4 $T$, 2 $T$,  1 $T$, 0.5 $T$ from the top curve to the bottom
one at the same temperature T = 95 K, (b) T = 95 K, 90 K , 80 K,
60 K, 40 K from the top to the bottom with H = 4 T.  }
       \label{figure7}
       \end{figure}

\begin{figure}
       \centering
       \includegraphics[width=8cm] {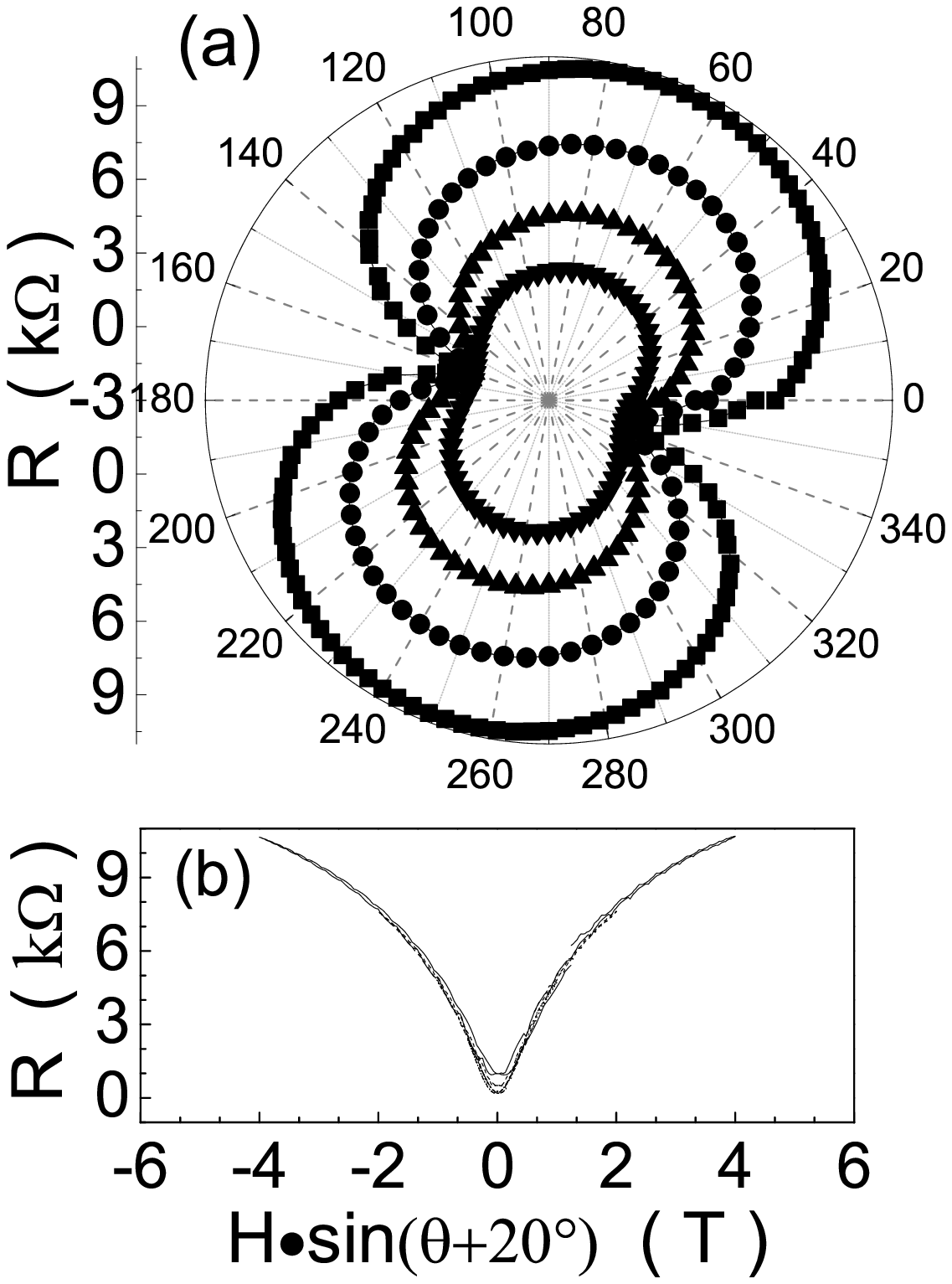}
       \caption{The angular and field dependence of resistance at
       fixed temperature ( $T$ = 95 $K$ ) for $S$-$t$. (a) $R$ vs angle $\theta$,
       H= 4 $T$, 2 $T$,  1 $T$, 0.5 $T$  from outside  to inside, (b) The scaling of $R$ in
       terms of H$\cdot\sin{(\theta+20^\circ)}$. }
       \label{figure8}
       \end{figure}

The issue of  anisotropy  is  related to the dimensionality which
is closely connected with the mechanism  of HTS, and some unusual
transport and magnetic properties have been observed and discussed
earlier\cite{IYE-comments,woo-prl,schmitt-prl,kes-prl} in the
strongly anisotropic systems. In these highly anisotropic HTS
systems, the Josephson coupling is weak, so the PVS in the
adjacent $ab$-planes are weakly coupled and the transport and
magnetic properties are nearly two dimensional\cite{clem-sst}.
Therefore, the 2$D$ scaling of our resistive data suggests the
thermally-activated dissipation mainly stems from the motion of
the PVS in the whole angular range in mixed state for both case of
$\alpha$ and $\beta$, which indicates that the two dimensional
behavior and the intrinsic anisotropy of the $Tl$-2212 films are
very strong.

For the thin film of $S$-$t$, we have also measured the resistance
as a function of $\theta$ under various fields at different fixed
temperatures with $\theta$  changed from 0$^{\circ}$ to
360$^{\circ}$ continuously.  The field dependence of resistance as
a function of angle $\theta$ is shown  in Fig. 6 with (a) T = 60 K
, (b) T = 80 K and in Fig. 7(a) with T = 95 K. In each panel, the
applied field H is 4 T , 2 T, 1 T, 0.5 T from the top curve to the
bottom one.  Under the same field H = 4 T, the resistance as a
function of angle $\theta$ at various  temperatures ( 95 K, 90 K,
80 K ,60 K, 40 K ) is shown in Fig. 7(b).   The resistivity of
$S$-$t$ is three or four magnitude of order larger than that of
$S$-$c$, which indicates that the resistance of sample $S$-$t$ is
mainly controlled by the resistivity along the c-axis. And the
$c$-axis resistive dissipation of $La_{2-x}Sr_xCuO_4$ has been
reported previously\cite{slyuan}. Clearly, the symmetric axis of
all curves is at $\theta \approx 70^\circ$ ( or 160$^\circ$ ),
which proves that the $ab$-plane of $S$-$t$ is truly $20^\circ$
tilted with respect to the surface of the substrate. This periodic
changing of resistance can be displayed obviously by drawing in
the polar coordinates, as shown in Fig. 8(a) with T = 95 K. In
this sense, the resistance measurement in this way  offers a
complementary method to determine the growth orientation of high
anisotropic HTS, especially for the thin films.

Furthermore, it is found that the resistance at the same
temperature can be also scaled in terms of $c$-axis-component
field $H\cdot\sin\beta$ ( $\beta=\theta+20^\circ$ ), which is
obviously seen from the scaled results at T = 95 K, as shown in
the Fig. 8(b). The scaling is poor at the angle
$\theta=160^\circ$, but the scaling has a high quality in wide
angle region.

The possible explanation of this scaling is the anisotropy of the
upper critical field. In present configuration, assuming that the
resistivity can be written as
\begin{equation}
 \rho=\rho_0 \cdot f[\frac{H(\beta)}{H_{c2}(\beta)}]
\end{equation}
where $H_{c2}(\beta)$ is the effective upper critical field. In
the anisotropic Ginzburg-Landau model, $H_{c2}(\beta)$ for an
arbitrary field angle is expressed by\cite{IYE-comments}
\begin{equation}
 H_{c2}(\beta)=\frac{H_{c2}^{\|c}}{\sqrt{\sin^2\beta+\cos^2\beta/\gamma^2}}
\end{equation}
Here, $\beta$ denotes the field angle defined from $ab$-plane. For
the two dimensional systems, this function reduces to
$H_{c2}(\beta)={H_{c2}^{\|c}}/{{\sin\beta}}$, and
$H^{\|ab}_{c2}>>H^{\|c}_{c2}$. As a result, $\rho=\rho_0 \cdot
f[\frac{H\cdot sin{\beta)}}{{H_{c2}^{\|c}}}]$. In this case, the
in-plane component field $H\cdot cos\beta$ is not enough to
destroy the coupling of CuO layers, and the resistance is only
dependent on the $c$-axis component of the applied field.

\section{Conclusions}
In conclusion, we have measured the resistance as a function of
angle between the applied field and $ab$-plane of the sample when
the angle is changed continuously at various  temperatures for a
$c$-axis-oriented and a tilted  $Tl$-2212 thin films. The
resistance with the changed angle demonstrates the special
symmetry structure of the sample. Moreover, it is found that the
resistance at the same fixed temperature can be well scaled  in
terms of the $c$-axis component of the applied magnetic fields.
Meanwhile, the measurement of resistance in this way can be
regarded as a complementary method to determine the growth
orientation of the high anisotropic HTS, which is specially
helpful for the tilted-growth thin film samples.

\section*{Acknowledgments}
This work is supported by the National Science Foundation of
China, the Ministry of Science and Technology of China, and the
Chinese Academy of Sciences within the knowledge innovation
project.

\end{document}